\documentstyle[prc,aps,preprint,epsfig]{revtex}

\begin{document}

\date{\today}

\title{Coulomb versus nuclear break-up of $^{11}$Be halo 
nucleus in a non 
perturbative framework.}
\author{M.Fallot$^{a)}$, J.A.Scarpaci$^{a)}$,  
D.Lacroix$^{b)}$, 
Ph.Chomaz$^{c)}$ and J.Margueron$^{a)}$}
\address{$^{a)}$ Institut de Physique Nucl\'eaire, 
IN2P3-CNRS, 91406 Orsay, France  \\  
$^{b)}$ LPC/ISMRA, Blvd du Mar\'{e}chal Juin, 14050\ 
Caen, France\\  
$^{c)}$ {\it G.A.N.I.L., B.P. 5027, F-14076 Caen Cedex 
5, France} \\
\begin{abstract}
The $^{11}$Be break-up is calculated at 41 MeV per nucleon incident energy
on different targets using a non perturbative time-dependent quantum
calculation. The evolution of the neutron halo wave function shows an
emission of neutron at large angles for grazing impact parameters and at
forward angles for large impact parameters. The neutron angular distribution
is deduced for the different targets and compared to experimental data. We
emphasize the diversity of diffraction mechanisms, in particular we discuss
the interplay of the nuclear effects such as the towing mode and the Coulomb
break-up. A good agreement is found with experimental data.
\end{abstract}
}

\maketitle

                                       


\section{introduction}

A major discovery of the last decade in nuclear physics is the
observation of halo nuclei\cite{Tan85}. The presence of these extended systems
has been uncovered by break-up studies. However, an unambiguous
interpretation of these reactions requires a deep understanding of reaction
mechanisms such as diffractive effects \cite{Han95}. In particular, the
interplay between nuclear and Coulomb dissociation is of major importance.

On the experimental side, measurements of the neutron angular distribution
have been performed for the one-neutron break-up of $^{11}$Be on Au, Ti and
Be targets \cite{Ann94} at 41 MeV per nucleon. They clearly present two
contributions. One peaked at small angles (below 10 degrees) which strongly
varies with the target ; for the Au target the cross section is as high as
50 barns per steradian at small angles whereas for the Be target it is
around 1 barn per steradian. A second one located at large angles which
shows a lesser dependence with the target ; the cross section around 30
degrees is about 100 mbarn per steradian for all studied targets. The small
angle region was understood through models based on Coulomb-excitation
theory to come from Coulomb dissociation\cite{Ann94} and the large angle
emission was thought to be a consequence of nuclear break-up. Perturbative
calculations\cite{Bon98-1,Bon98} including both interactions or non
perturbative calculations\cite{mel99,typ99} which only account for the
Coulomb field have been performed in the last years. Only recently the time
dependent Schr\"odinger equation was solved \cite{Lac99} and applied to the
break-up of $^{11}$Be using different numerical techniques than the one we
have used \cite{esb01}. In this later paper, the authors predict the
presence of neutrons emitted with a high angular momentum in the $^{11}$Be
frame which corresponds to the neutrons measured at large angles that were
mentioned above. This prediction, which the authors do not observe when they
use the eikonal approximation, is in agreement with our calculation as it
will be discussed further in this paper. At the same time our paper was in
submission, we noted the work of ref.\cite{typ01} that also presents a method
of solving the time dependent Schr\"odinger equation and applies it to the
break-up of $^{11}$Be and $^{19}$C to calculate the relative energy between
the emitted neutron and the core.

In our article, we present a non-perturbative model which accounts both for
the Coulomb and the nuclear effects through the resolution of the full
time-dependent Schr\"odinger equation onto a cartesian mesh.
Our calculation shows that the emission of neutrons at large angles is due
to the interaction of the particle with the nuclear potential. Such an
emission has already been observed in reactions between stable nuclei
\cite{Sca98} where the nucleon originated from the target and not the
ejectile as in the case of $^{11}$Be break-up. This mechanism was called
"towing mode" as the particle was pulled out from the target and towed by
the projectile for a short while. Through the resolution of the
time-dependent Schr\"odinger equation we reproduced this emission to the
continuum with specific angles and energies in agreement with the
measurements \cite{Lac99}.

In the following, we will show through our calculation that both Coulomb and
nuclear fields play an important role in the case of the neutron break-up of
halo nuclei.

\section{Description and inputs of the model}

We use the calculation presented in a previous paper \cite{Lac99} and apply
it to the reactions $^{197}$Au, $^{48}$Ti, $^{9}$Be ($^{11}$Be, $^{10}$Be +
n). This calculation describes the wave function distortion of an initially
bound particle as it passes by the potential induced by the reaction
partner. This is performed in the framework of independent particles.

The wave function of the neutron halo is deduced from the potential found
by N. Vinh Mau \cite{Vin95} to reproduce the inversion of the 2s and the 1p
states in the $^{11}$Be nucleus for which a derivative of the Wood-Saxon
potential is added at the surface. The potential between the neutron and the 
$^{10}$Be core reads
\begin{eqnarray}
V_{Be} (r) = 
\frac{V_0} {1+ e^{\frac{r-R}{a}}} + 16. \alpha  
\frac{e^{2\frac{r-R}{a}}}{\left(1+ 
e^{\frac{r-R}{a}}\right)^4}
\label{Eq:1}
\end{eqnarray}
where the diffuseness $a$ is 0.75 fm and the coefficient $\alpha$ is equal
to -10.56 MeV for the 2s state. The radius R equals to $1.27\times A^{1/3}$ fm.
Numerically the initial wave-packet is obtained by diagonalizing the
one-body Hamiltonian in spherical coordinates in a sphere of 30 fm radius
with a space-step of $\Delta r$ =0.02 fm. The depth $V_0$ was taken to be
-40 MeV. The wave function calculated in spherical coordinates for 
$^{11}$Be is then mapped onto Cartesian coordinates in order to calculate
the dynamical
evolution. Special attention has been given to the purity of the ground
state $^{11}$Be neutron halo by performing an additional imaginary time
evolution on the Cartesian network, after which the 2s state is 
bound with -0.503 MeV.

The dynamical evolution of the system is given by the following single
particle Schr\"odinger equation which reads in the $r$ space
\begin{eqnarray}
i\hbar \frac{d}{dt} \varphi({\bf r}) = 
\left(\frac{-\hbar^2}{2m}\Delta + V_T\left({\bf r}-{\bf r}_T(t)\right)
+ V_{Be}\left({\bf r}-{\bf r}_{Be}(t)\right) \right) 
\varphi({\bf r})
\label{Eq:2}
\end{eqnarray}
where $V_{T}$ and $V_{Be}$ are the time-dependent potentials between the
neutron and the target and between the neutron and the projectile
respectively. ${\bf r}_{T}(t)$ and ${\bf r}_{Be}(t)$ correspond to the
target and the projectile positions respectively. The nuclear potential of
the Au and Be target, $V_{T}$(r), is taken to be of Wood-Saxon shape with a
diffuseness $a=0.5$ fm, a radius $R=1.27 A^{1/3}$ fm ($A$ being the mass
number of the $^{197}$Au or $^9$Be) and of depth adjusted to obtain the
experimental binding energy of the last neutron. For the Au target, $V_0$ =
-49.3 MeV, giving a 3p state bound by about 8 MeV and for the Be target,
$V_0$ = -40. MeV, giving a 1p state bound by about 1.66 MeV. For the
$^{48}Ti$ target, we used the Becchetti and Greenlees \cite{bec69}
prescriptions to obtain the real part of the nuclear potential between the
neutron and the Ti, leading to a depth of $V_0$ = -41.2 MeV, a diffuseness
of 0.75 fm and a r$_0$ value of 1.17 fm.

In order to take into account the target and the $^{10}$Be core
displacement, the evolution is performed for each time step using the
classical Coulomb trajectory for the center of mass motions (${\bf r}_T(t)$,
${\bf \dot{r}}_T(t)$, ${\bf r}_{Be}(t)$, ${\bf \dot{r}}_{Be}(t)$) from a
distance of -400 fm along the initial velocity axis between the projectile
and the target and equal to the impact parameter $b$ on the perpendicular
axis. This explicit treatment of the core recoil is responsible for the
Coulomb excitations of the neutron around the $^{10}Be$ core. The numerical
method used for the trajectory calculation is the Runge-Kutta method.

We have chosen to use the split-operator method \cite{Fei82} for the time
evolution. This is a well known way to solve the time dependent
Schr\"odinger equation (Eq.\ref{Eq:2}) on a 3D cartesian lattice. It is
faster and as accurate as the methods based on Taylor expansion of the
evolution operator which are also routinely used to solve time dependent
problems. Moreover it allows to treat large lattice performing a rather long
time calculation. 

The time step used is of 1.7 fm/c on a mesh of 64x64x64 fm$^3$ with a step
size of 0.5 fm. We have tested the numerical accuracy of the method with
respect to the mesh parameters. The calculation is performed in the initial
projectile frame and a Galilean frame transformation is performed to extract
the observed quantities.

\section{Results}

\subsection{Density plots}

\begin{figure}[tbph]
\begin{center}
\epsfig{file=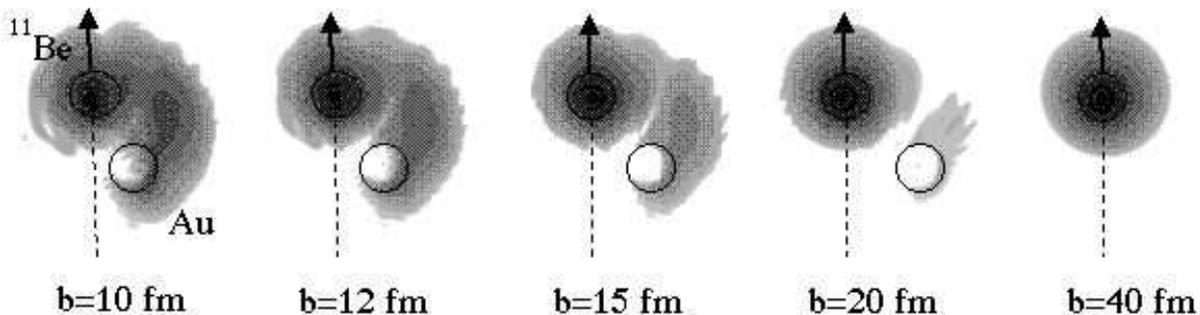,width=16.cm}
\end{center}
\caption{
Density plots of the 2s wave function in $^{11}$Be after scattering on a
$^{197}$Au target for impact parameters of 10, 12, 15, 20 and 40 fm. The
Coulomb trajectory is represented by the dashed line for each impact
parameter. These plots are displayed in a logarithmic scale,
black representing the most intense density. The ratio in density
is 3 between each grey area. }
\label{fig:1}
\end{figure}

The result of these calculations (Fig.\ref{fig:1}) is presented for the
$^{11}$Be + Au reaction as the probability density integrated
over the z-axis, $\rho_{2s}(x,y)=\int\phi(x,y,z)^2dz$, for impact parameters
b = 10, 12, 15, 20 and 40 fm, after the evolution of the 2s wave function
initially in the $^{10}$Be potential.

After the evolution, we subtract the contributions of the wave-functions
which are bound in the $^{10}$Be potential in order to keep only the emitted
part of the wave-function. This subtraction is performed by projecting out
all the bound states $|\alpha>$ of the neutron in the $^{10}$Be, leading to the emitted wave function $|\psi>$.
\begin{eqnarray}
|\psi> = | \phi> - \sum_{\alpha} | \alpha> 
<\alpha|\phi> 
\label{Eq:3}
\end{eqnarray}
The bound wave functions in the $^{10}$Be core potential are the 2s state
itself bound by -0.503 MeV, the 1s state bound by -23 MeV and the 1p state
using the same nuclear potential and which is then bound by about -11 MeV.
Note that we do not consider the contribution of the 1p wave function which
has been experimentally evidenced at -0.183 MeV since it is not a bound wave
function of our potential.

\begin{figure}[tbph]
\begin{center}
\epsfig{file=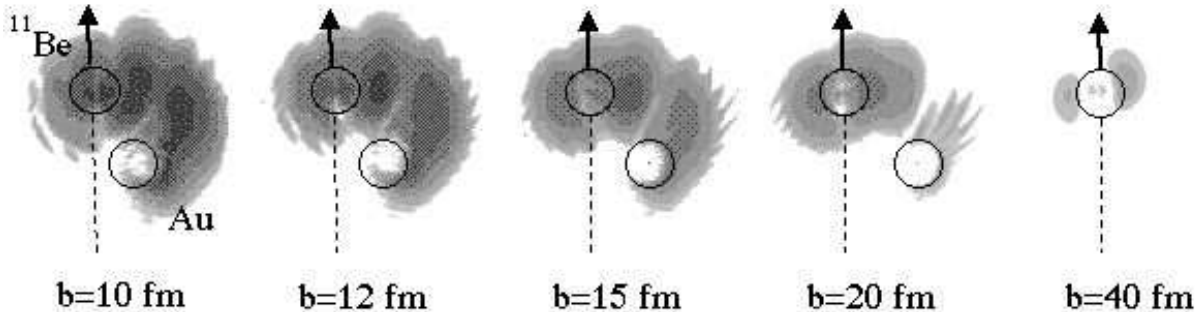,width=16.cm}
\end{center}
\caption{
Density plots of the 2s wave function of $^{11}$Be after scattering on a
$^{197}$Au target with an impact parameter of 10, 12, 15, 20 and 40 fm and
after subtraction of the bound wave functions. }
\label{fig:2}
\end{figure}

Fig.\ref{fig:2} shows the same evolutions as Fig.\ref{fig:1} after the
subtraction of these components. A sizeable fraction of the wave function is
then removed around the $^{10}$Be core position. 
One can see that the small impact parameters (b $\le$ 20 fm) are responsible
for neutrons emitted at large angles in the opposite direction of the core
compared to the initial direction. For larger impact parameters, neutrons are
forward focussed. This small angle emission can be understood as a Coulomb
dissociation in which the $^{10}$Be core has been shaken by the Coulomb
field of the target. In our calculation, for large impact parameters, we
observe a displacement of the neutron wave function compared to the core
which might be the first oscillation of a soft dipole resonance as it has
been suggested in ref.\cite{nak94}. Due to the weak binding of the last
neutron, the halo has separated from the core, transferring little momentum
to the neutron which is then emitted along the initial trajectory. 
The calculation shows a weak emission (see Fig.\ref{fig:2} right) which must,
however, be integrated over a large impact parameter domain to obtain the
total cross section.

In order to compute cross-sections, we used the following method. For large
impact parameters the fraction of wave function emitted is small and thus
according to the perturbation theory, to gain time and reduce the error
inherent to the method, we performed the calculation up to the minimum
distance of approach and multiplied the result by two. For large impact
parameters we have tested that this procedure gives the same results as the
complete calculation demonstrating that the impact parameters are large
enough to guarantee a first order perturbation.

On the contrary, for small impact parameters (below 20 fm) the nuclear
break-up cross section is large and the nuclear potential spreads the wave
function on the whole mesh (see Fig.\ref{fig:1} and \ref{fig:2}) we stopped
the calculation at a distance of 20 fm after the projectile has passed by
the target and checked that the extracted values do not change for a slightly
longer evolution. Since the nuclear break-up is much larger than the Coulomb
break-up cross section for the small impact parameters, as will be shown in
section B, the fraction of Coulomb break-up missed there is negligible.

\subsection{Interplay between nuclear and Coulomb break-up.}

\begin{figure}[tbph]
\begin{center}
\epsfig{file=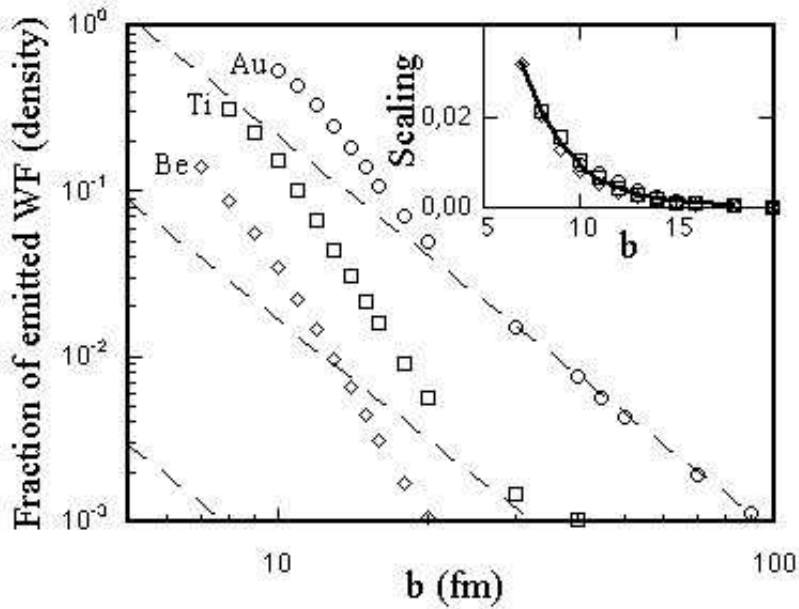,width=11.cm}
\end{center}
\caption{Fraction of wave function emitted (${\rm\sl Frac}$) versus the
impact parameter for the three targets (symbols). Dashed lines are fits
obtained with the formula $\alpha b^{\beta}$ for the large impact
parameters (see text). Insert: scaling of ${\rm\sl Frac}-\alpha b^{\beta}$ with
$A_{T}^{2/3}$ (symbols). The solid line is the fit of the scaling with 
$e^{-\gamma b}$ (see text).}
\label{fig:3}
\end{figure}

Since the $^{10}$Be core is detected in the experiment, we assumed a minimum impact
parameter corresponding to a grazing trajectory. We followed the strong absorption
model for which the minimum parameter is calculated as
1.4($A_{T}^{1/3}+A_{Be}^{1/3}$) fm yielding to $b_{min}=$ 11 fm for the Au
target, $b_{min}$= 8 fm for the Ti target and $b_{min}=$ 6 fm for the Be
target. To obtain a total cross section, this calculation was performed for
impact parameters running from the grazing parameter, to an impact parameter
where the fraction of wave function emitted no longer changes with increasing
impact parameter. This is shown in Fig.\ref{fig:3} where the
fraction of wave function emitted ${\rm\sl Frac}= \int d^3r|\psi|^2$ is plotted 
versus the impact parameter.

For the Au target this percentage starts to saturate around 200 fm 
whereas for the Ti
it is around 100 fm and 40 fm for the Be target. At these impact parameters
the fraction of wave function emitted is around 0.04$\%$. At the
saturation, only the noise inherent to the numerical method remains and we
do not include larger impact parameters in the cross section estimate.

In this figure we clearly see a change in the slope of ${\rm\sl Frac}$, between the
small and the large impact parameters which could indicate the transition
between the nuclear and the Coulomb perturbation.

In order to get a deeper insight in the interplay between nuclear and
Coulomb break-up, we have studied the mass and charge dependence of the
fraction emitted (${\rm\sl Frac}$) with the impact parameter. At large impact
parameter, above 40 fm, the wave function does not see the target nuclear
potential and only the Coulomb field is felt by the core. We observe in
Fig.\ref{fig:3} that at large impact parameters ${\rm\sl Frac}$ can be fitted with
the formula $\alpha b^{\beta}$ for the Ti and Au targets. We found
$\beta=-2.4$ and $\alpha$ which scales quadratically with the charge as
$\alpha=0.0088 \times Z^2$. Fits are shown on Fig.\ref{fig:3} as dashed
lines (we assumed the same dependence for the Be target). Since the
probability of emission is small we can conclude that the excitation is
perturbative. This is confirmed by the $Z^2$ dependence of this probability.

We extrapolated the Coulomb effect to the small impact parameters and by
subtracting the fit curve to the fraction emitted we obtained a function
which we expect to only contain the nuclear effect. Indeed, after
renormalizing these results with $A_T^{2/3}$, we obtained a quantity named
$Scaling = (Frac - \alpha b^{\beta})/A_T^{2/3}$, which no longer depends on
the target (see insert of Fig.\ref{fig:3}). This leads to a cross section
that scales with $A_T^{2/3}$ as expected for the nuclear emission
\cite{kob89}. This curve can be fitted in turn by an exponential
(Scaling $\propto e^{-\gamma b}$ with $\gamma = 0.40 \pm 0.01$ fm$^{-1}$).

We would like to emphasize that this empirical fitting method indicates a
rather good separation between Coulomb and nuclear effects in our
calculation. Such a separation, which comes as an hypothesis in perturbative
framework, is however not natural in our framework where both effects are
accounted for at the same time and can therefore interfere. By treating the
Coulomb and nuclear part separately we have controlled that the possible
interferences are small.

\begin{figure}[tbph]
\begin{center}
\epsfig{file=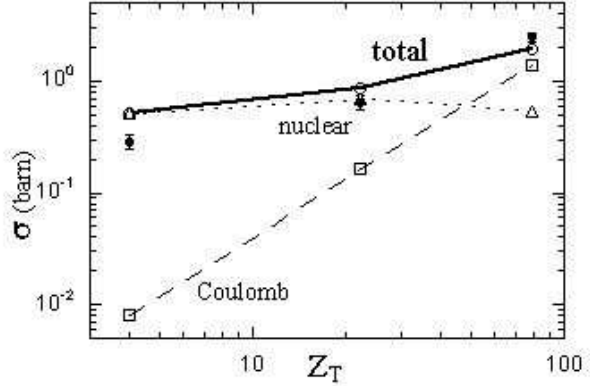,width=8.cm}
\end{center}
\caption{Coulomb (squares) and nuclear break-up (triangles) cross sections as a
function of the target charge ($Z_{T}$) obtained from the fits to the
calculated fraction of the wave function emitted and the fit to the Scaling
parameter, integrating from the corresponding $b_{min}$. The Coulomb
contribution scales with $Z^{1.65}$ (not shown in the figure). The sum of
the Coulomb and the nuclear cross-section is also reported as circles joined
by a plain line and compared to the experimental data (plain circles with error
bars) of ref. [3].}
\label{fig:3bis}
\end{figure}

Using the parameters extracted from the fits and integrating over the impact
parameter, we extract the variation of the break-up cross section with the
charge of the target for both the Coulomb and the nuclear interactions. This
evolution is presented in Fig.\ref{fig:3bis}. We observe a Coulomb cross
section that scales with $Z^{1.65}$ close to the value of 1.725 from
ref.\cite{han87}.

It should be noticed that in the case of the Au target we have checked this
analysis by turning on and off the nuclear and the Coulomb field separately in
the calculation.

\subsection{Cross sections and angular distributions}

Once we have the fraction of wave-function emitted we can access the
differential cross-section by integrating the corresponding probability 
between $b_{min}$
and $b_{max}$ defined in previous section. For each impact parameter $b_n$
(fm), the cross section is equal to $2\pi b_n \Delta b_n.10^{-2}$ (barn)
times the modulus squared of the fraction of wave function which is emitted, where
$\Delta b_{n}$ is taken to be equal to $(b_{n+1}-b_{n-1})/2$. The step in
impact parameter between two calculations varied from 1 fm (for small impact
parameters) to 40 fm above $b=150$ fm.

\begin{figure}[tbph]
\begin{center}
\epsfig{file=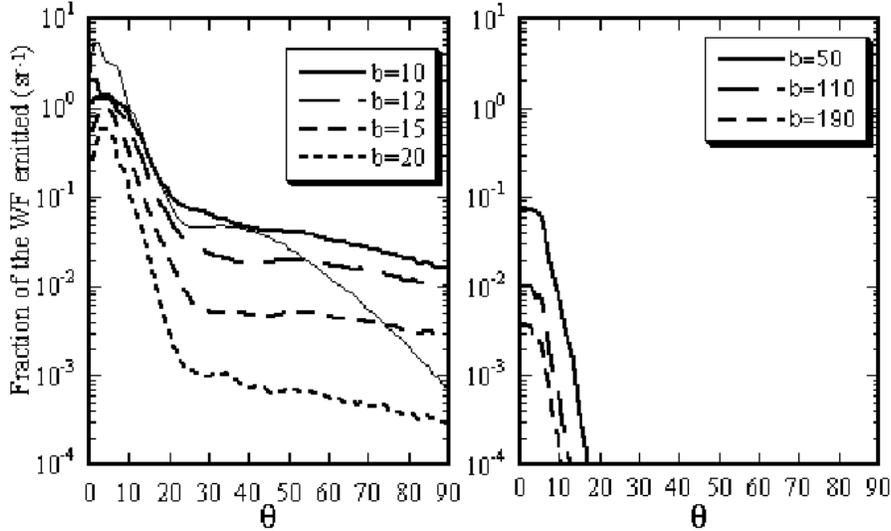,width=12.cm}
\end{center}                         
\caption{
Angular distributions of the fraction of the wave function emitted for
impact parameters from 10 to 20 fm (left) and from 50 to 190 fm (right). On
the left figure the Fourier transform of the initial 2s wave function is
also shown as a plain line. }
\label{fig:4}
\end{figure}

The angular distribution of the emitted neutron is extracted by applying the
Fourier transform to obtain the remnant part of the wave function in the
momentum space \cite{Lac99} (Fig.\ref{fig:4}). 
\begin{eqnarray}
\rho({\bf p}) = \left|\int exp(-i{\bf p}{\bf r}/\hbar) \Psi ({\bf r}) d{\bf r}\right|^2
\end{eqnarray}
It should be noticed that to fully take into account the final state
interaction this calculation should be performed on the asymptotic wave
function. We have controlled that $\rho({\bf p})$ does not evolve anymore for
longer evolution. Therefore, it can be considered as the final momentum
distribution of the emitted particles. For small impact parameters, nuclear
effects are important and the nuclear refraction of the neutron halo gives
rise to an emission at large angles. This can be seen in the spectra
extracted from the calculations performed at impact parameters of 10 to 20
fm which exhibit a component at angles above 30 degrees. This component is
also seen in the density plots of Fig.\ref{fig:2} and present an
anti-correlation with the core trajectory compared to the incident
direction. This emission to the continuum is known as Towing Mode. In the
case described with stable nuclei, a nucleon from the target is pulled out,
towed along by the projectile and finally expelled at large angles and large
velocities in the laboratory frame \cite{Sca98}. In the case of the
$^{11}$Be experiments, the emitted neutron belongs to the projectile (the
$^{11}$Be) which breaks up as it is perturbed by the target nuclear
potential. In the $^{11}$Be frame, this nuclear break-up can be seen as the
emitted neutron being accelerated by the target potential (see
Fig.\ref{fig:2}). We expect the same mechanism to be present for this halo
nucleus as for a stable nucleus. However, in the case of a halo nucleus, the
cross section of this nuclear break-up should be increased compared to a
stable nucleus, due to the large extension of the neutron wave function and
thus the larger impact parameter at which it sees the nuclear potential of
the other nucleus. This will be discussed further when comparing with a more
bound wave function (see section D). We also display the angular
distribution of the initial 2s wave function. In a sudden core removal model
it would correspond to the distribution of the emitted neutrons. However it
differs from the presented distributions as well as from the experimental
distribution of ref.\cite{Ann94}, showing the importance of the reaction
mechanism in the break-up process.

At larger impact parameters, above 50 fm, only the component between 0 and
15 degrees remains (Fig.\ref{fig:4} right). There, the neutron wave function
does not feel the nuclear potential anymore and the break-up comes from the
deviation of the $^{10}$Be core in the Coulomb field. Indeed, because the
halo neutron of $^{11}$Be is so weakly bound, a light shaking of the core is
enough to induce the break-up. This, in turn, is not expected with a more strongly
bound nucleus as it will be discussed further in section D.

\begin{figure}[tbph]
\begin{center}
\epsfig{file=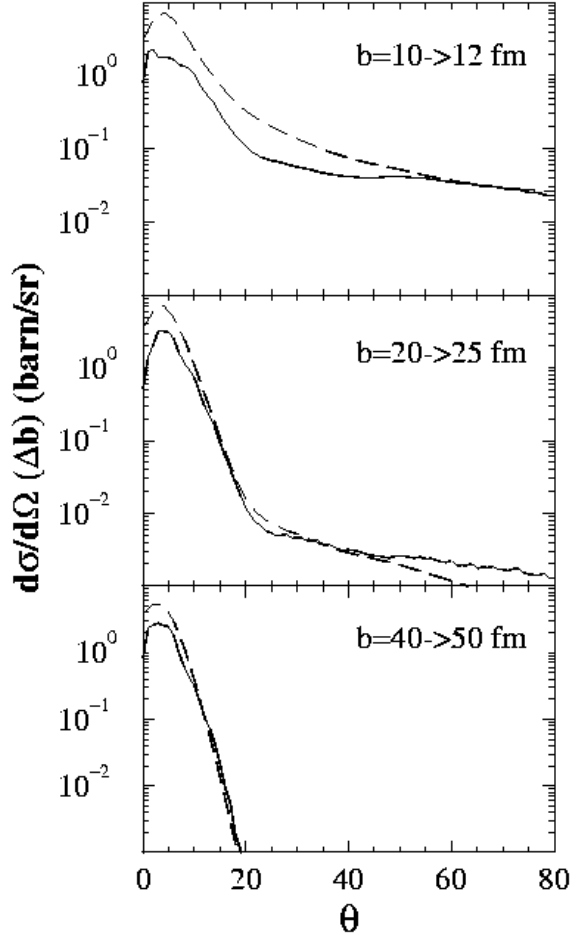,width=8.cm}
\end{center}
\caption{
Angular distribution for the emission of the $^{11}$Be halo neutron using a
perturbative approach (dashed lines) and our calculation (plain lines) for
three regions of impact parameter.}
\label{fig:6}
\end{figure}

We have compared our calculation with a time dependent perturbative approach
proposed by Bonaccorso, Brink and Margueron presented in
ref.\cite{Bon98,Bon00} (called here perturbative) where they treat both
the Coulomb and the nuclear break-up as a transfer to the continuum. The
result we show have been compared with the same optical potential (presented
here) and the wave function of the
neutron halo is taken as the outer part of a 2s wave-function and normalized
to the wave function used in our dynamical evolution. At small impact
parameters we know that a large fraction of the wave function is emitted
(Fig.\ref{fig:3} shows that more than 50\% of the wave function is emitted
for 10 fm of impact parameter) which indicates that the perturbative approach
may not be adequate in the case of the Au target, however at large impact
parameters this value becomes small (1\% at 30 fm of impact parameter) and
there the perturbative approach is well justified. We thus expect the
results of our calculations to be different from the perturbative approach
for small impact parameters and to become more and more similar as we go at
larger impact parameters. Fig.\ref{fig:6} shows this comparison for three
regions of impact parameters for the break-up on a Au target.

For the region between 10 and 12 fm the two calculations predict a quite
different shape of the neutron angular distribution. At large impact
parameter, the two calculations exhibit similar shapes of the neutron
angular distributions but a larger cross section is observed around 5
degrees in the perturbative approach. This difference, present at all impact
parameters, might be due to the remaining differences between the
ingredients of the two methods. In particular it should be stressed that the
wave functions used are not the same in both cases. In the perturbative
calculation the inner part of the wave function is the Hankel function h$_1$
while in our approach we use the exact solution of time
dependent Schr\"odinger equation. We compared two time dependent theories
which show some discrepancies. The use of these theories as a spectroscopic
tool for halo nuclei requires a better understanding of these differences. A
deeper analysis is in progress and will be presented in a forthcoming
article. 

\subsection{Comparison between halo and non-halo neutron emissions}

\begin{figure}[tbph]
\begin{center}
\epsfig{file=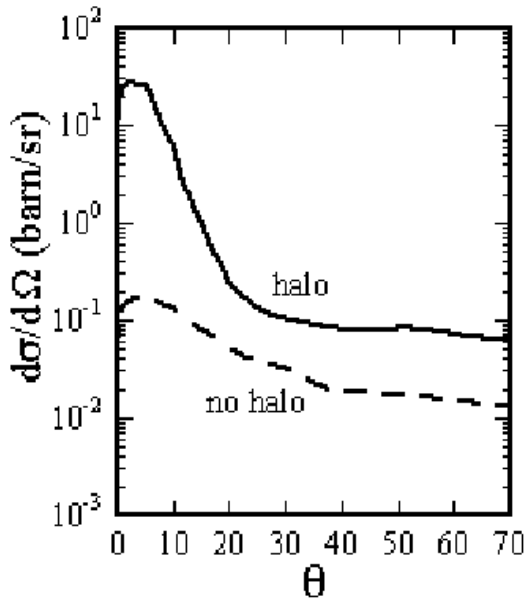,width=7.cm}
\end{center}
\caption{
Angular distribution for the emission of the $^{11}$Be halo neutron (plain
line) and a non-halo neutron (dashed line) bound by 7 MeV after the
scattering on a Au target summed over impact parameters between 10 fm and
110 fm. }
\label{fig:7}
\end{figure}

Our calculation has also been performed to infer the evolution of a strongly bound
wave function. We used a Wood-Saxon potential with $V_0$=-70.5 MeV, A=11,
and obtained a 2s wave function bound by 7 MeV. The calculation has been
performed for a Au target and for impact parameters running from 10 fm to
110 fm. The result is shown in Fig.\ref{fig:7} and is compared to the
break-up of the $^{11}$Be halo for the same impact parameter range. The
differential cross
section of neutron emission for the bound nucleus is more than hundred times lower than for $^{11}$Be
below 10 degrees. For large angles, around 40 degrees, the differential
cross section is about 4 times lower. This latter difference can be understood by the
extension of the wave function, which is much larger in case of a halo.

In particular, it is often argued that Coulomb effects shadow the nuclear
mechanism in heavy targets like Au. Although the relative proportion of
these two effects is largely in favor of Coulomb dissociation for a heavy
target such as Au, we show that nuclear break-up is of major importance for
large angle emission with a cross section around 0.5 barn. Whereas Coulomb
dissociation on a heavy target such as Au could be a direct measurement of
the binding energy of the particle, the amount of towed particle could in
turn bring information on the extension of the wave function, and hence
answer the question whether we are dealing with a halo state or not.
Furthermore, in stable nuclei, it has been shown that transfer to the
continuum due to the Towing Mode might be a tool to infer information on
shell structure as presented in ref.
\cite{Sca98}. One might then be able to use this large angle emission to
obtain additional information on nuclear halo wave-function properties.

\subsection{Comparison with the $^{11}$Be data}

\begin{figure}[tbph]
\begin{center}
\epsfig{file=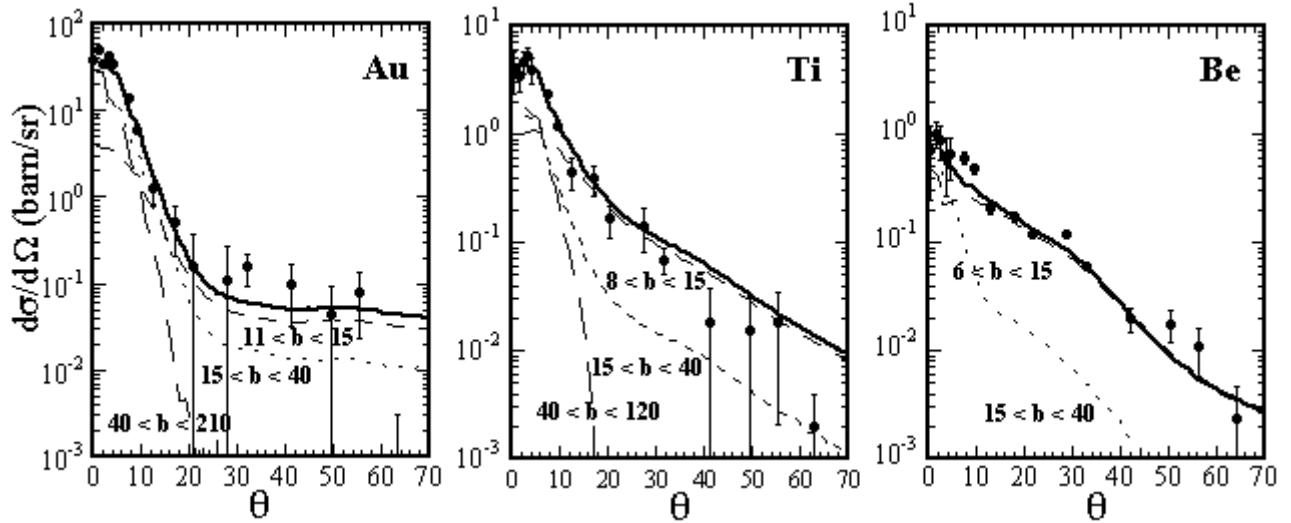,height=7.cm}
\end{center}
\caption{
Calculated angular distributions for impact parameter running from $b_{min}$
to $b_{max}$ (11 to 210 fm for the Au target (left figure)) (plain line) for
neutron of energy higher than 26 MeV. Dots with error bars are data points
from ref.[3]. Middle figure is for the Ti target and right
figure for the $^{9}$Be target. Contribution of the calculation for three
impact parameter regions, $b_{min}$ to 15 fm (short dashed lines), 15 to 40 fm
(dotted lines) and 40 to $b_{max}$ (long dashed lines) are also presented. }
\label{fig:5}
\end{figure}

To compare our calculation to the data of ref. \cite{Ann94}, we extracted
the differential cross section by summing the calculations from $b_{min}$ up
to $b_{max}$. For the Au target we took $b_{max}$ = 210 fm, 120 fm for Ti
and 40 fm for Be. Calculations are shown in Fig.\ref{fig:5}, multiplied by
0.84 to take into account the spectroscopic factor of the 2s state found
experimentally and reported in ref.\cite{For99}. The calculation is compared
to the experimental data of ref.\cite{Ann94} for all three targets taking
into account the experimental threshold of 26 MeV for the neutron detection.
Note that our calculation includes all the inelastic channels in which the
target is excited but also those in which the $^{10}$Be remnant is in an
excited state below its neutron separation energy. This is also included in
the data that measured the $^{10}$Be and the neutron in coincidence. However
our calculation does not take into account the possible two body dissipation
since we use a single-particle framework. The calculations (plain lines) are
in good agreement with the experimental data, both at small and large
angles. In Fig.\ref{fig:5} we have reported the result of the calculations
for three impact parameter regions showing the separation between the
nuclear and the Coulomb break-up. The simultaneous reproduction of the data
for the Au the Ti and the Be targets demonstrates that the Coulomb and the
nuclear interactions are well taken into account. In these calculations the
contribution of a neutron in a 1d state coupled to a $2+$ excitation of the
core as deduced from the experiment
\cite{For99} has not been taken into account so far. However, since we already
reproduce the whole cross section for the Au target with 84\% of the 2s
break-up, the contribution of the 1d state would seem to be small in this
case.

\subsection{Relative energy}
\begin{figure}[tbph]
\begin{center}
\epsfig{file=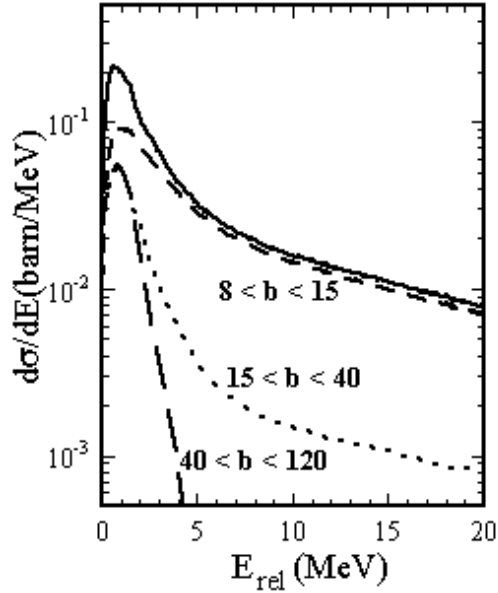,height=8.cm}
\end{center}
\caption{
Relative energy between the neutron and the Be core after breakup on a Ti
target for impact parameter running from $b_{min}$=8 fm
to $b_{max}$=120 fm (plain line). Contribution of the calculation for three
impact parameter regions, 8 to 15 fm (short dashed lines), 15 to 40 fm
(dotted lines) and 40 to 120 fm (long dashed lines) are also presented. }
\label{fig:erel}
\end{figure}

We have calculated the relative energy between the emitted neutron and the
$^{10}$Be using the momentum of the core after the Coulomb trajectory. This
is presented in Fig.\ref{fig:erel} for three different impact parameter
regions for the Ti target. This spectrum shows that large
relative energies correspond to small impact parameter hence the nuclear
break-up whereas large impact parameters lead to small values of the relative
energy as expected for the Coulomb break-up. 

\section{ Conclusion}

We have investigated the neutron break-up of a halo nucleus in
the reactions Au, Ti, Be ($^{11}$Be, $^{10}$Be+n) at 41 MeV per nucleon, in
the framework of a time dependent quantum model. Results were compared with
the experimental neutron angular distributions of Ref.\cite{Ann94} and a
good agreement was found. Our calculation, that includes both the Coulomb
and the nuclear interactions, confirms that the forward peaked neutrons are
due to the Coulomb break-up and that the neutrons emitted at large angles
come from the interaction of the halo neutron with the target nuclear
potential. This diffractive mechanism is called towing mode
\cite{Sca98}. A strong angular correlation between the towed particle and
the projectile was observed in reactions between stable nuclei and is also
expected for the $^{11}$Be break-up. However the experiments performed so
far did not measure the scattering angle of the remnant $^{10}$Be. Our
calculations show that neutrons emitted below 15 degrees arise from the
Coulomb dissociation and most of the cross section comes from large impact
parameters as the neutron halo of $^{11}$Be is weakly bound. The shaking of
the $^{10}$Be core by the Coulomb field leads to the dissociation of the
halo and the emission of the neutron in the forward direction. This can also
be understood as a Coulomb excitation of $^{11}$Be above the particle
threshold, followed by neutron emission. This is the only mechanism
contributing to the break-up when the impact parameter is such that the halo
wave function does not overlap with the nuclear perturbative potential. The
Coulomb break-up is very much hindered for strongly bound neutrons, whereas
the nuclear break-up decreases by a factor of four due to the lesser
extension of the wave function. Calculations for the 1d wave function should
be performed for the $^{11}$Be break-up to infer its contribution and
understand better the data. More generally, those calculations could be used
to extract information on the wave function of the last bound neutron for
unstable nuclei for which the properties are not well known yet, provided
that a measurement of the neutron angular distribution cross section is
performed both at large and small angles.


\begin{references}
\bibitem{Tan85} {I. Tanihata et al, Phys. Rev. Lett. 55 (1985) 380. 
I.Tanihata, J. Phys. G22 (1996) 157.}
\bibitem{Han95} {P.G. Hansen, A.S. Jensen and B. Jonson, Annu. Rev. Nucl. 
Part. Sci. 45 (1995) 591.}
\bibitem{Ann94} R.Anne et al., Nucl. Phys. A575 (1994) 125.
\bibitem{Bon98-1} A.Bonaccorso, D. Brink Phys. Rev. C 57 (1998) 22.
\bibitem{Bon98} A.Bonaccorso, D. Brink Phys. Rev. C 58 (1998) 2864.
\bibitem{mel99} V.S.Melezhik and D.Baye Phys. Rev. C 59, (1999) 3232.
\bibitem{typ99} S.Typel and H.H.Wolter, Z.Naturforsh. 54A, (1999) 63.
\bibitem{Lac99} D.Lacroix, J.A.Scarpaci and Ph.Chomaz, Nucl. Phys. A 658 (1999) 273.
\bibitem{esb01} H.Esbensen and G.F.Bertsch, Phys. Rev. C 64, 014608
\bibitem{typ01} S.Typel and R. Shyam, Phys. Rev. C, to be published
\bibitem{Sca98} J.A.Scarpaci et al., Phys. Lett. B 428 (1998) 241.
\bibitem{Vin95} N.Vinh Mau, Nucl. Phys. A 592 (1995) 33.
\bibitem{bec69} F.D.Becchetti, JR. and G.W.Greenlees, Phys. Rev. vol 182 (1969) 1190
\bibitem{For99} S.Fortier et al., Phys. Lett. B461, 22 (1999).
\bibitem{Fei82}  M.D.Feit, J. Fleck, Jr. and A. Steiger, J. Comput. Phys. 
(1982) 412; M.D. Feit and J.A. Fleck, Jr., J. Chem. Phys. \textbf{78}
(1983) 301; ibid., \textbf{80} (1984) 2578.
\bibitem{nak94} T.Nakamura et al., Phys. Lett. B331, (1994), 296.
\bibitem{kob89} T.Kobayashi et al, Phys.Lett. B232 (1989) 51.
\bibitem{han87} P.G.Hansen and B.Jonson, Europhys.Lett. 4 (1987)409.
\bibitem{Bon00} J.Margueron, A.Bonaccorso and D.Brink, submitted to Nucl.
Phys. A.
\end{references}
\end{document}